\documentclass[epsfig,12pt,onecolumn]{article}
\usepackage{amsfonts}
\usepackage{amssymb}
\usepackage{amsmath}
\usepackage{multicol}
\usepackage{graphicx}
\usepackage{float}
\usepackage{caption}

\makeatletter \@addtoreset{equation}{section}

\def \be{\begin{equation}}
\def \ee{\end{equation}}
\def \bea{\begin{eqnarray}}
\def \eea{\end{eqnarray}}

\newcommand{\nc}{\newcommand}
\nc{\al}{\alpha} \nc{\bib}{\bibitem} \nc{\la}{\lambda}
\nc{\C}{\mbox{\hspace{1.24mm}\rule{0.2mm}{2.5mm}\hspace{-2.7mm} C}}
\nc{\R}{\mbox{\hspace{.04mm}\rule{0.2mm}{2.8mm}\hspace{-1.5mm} R}}
\input{tcilatex}
\begin{document}

\title{\textbf{Pulsar timing array results sheds light on Hubble tension
during the end of inflation}}
\author{M. Bousder$^{1}$\thanks{%
mostafa.bousder@fsr.um5.ac.ma}, E. Salmani$^{1,2}$, A.Riadsolh$^{3},$ \and %
H. Ez-Zahraouy$^{1,2}$, A. El Fatimy$^{4,5}$ and M. El Belkacemi$^{3}$ \\
$^{1}${\small Laboratory of Condensend Matter and Interdisplinary Sciences,
Department of physics,}\\
\ {\small Faculty of Sciences, Mohammed V University in Rabat, Morocco}\\
$^{2}${\small CNRST Labeled Research Unit (URL-CNRST), Morocco}\\
$^{3}${\small Laboratory of Conception and Systems (Electronics, Signals and
Informatics)}\\
\ {\small Faculty of Sciences, Mohammed V University in Rabat, Morocco}\\
$^{4}${\small Central European Institute od Technology,}\\
\ {\small CEITEC BUT, Purky\v{n}ova 656/123, 61200 Brno, Czech Republic}\\
$^{5}${\small Departement of Physics, Universit\'{e} Mohammed VI
Polytechnique, Ben Guerir 43150, Morocco}}
\maketitle

\begin{abstract}
Recently, pulsar timing array (PTA) collaborations, including NANOGrav, have
reported evidence of a stochastic gravitational wave background within the
nHz frequency range.\ It can be interpreted by gravitational waves from
preheating era. In this context, we demonstrate that the emission of this
stochastic gravitational wave background can be attributed to fluctuations
occurring at the end of inflation, thus giving rise to the Hubble tension
issue. At the onset of inflation, the value of the frequency of the
gravitational wave signal stood at $f=0.08nHz$, but it rapidly transitioned
to $f=1nHz$ precisely at the end of inflation. However, just before the end
of inflation, a phase characterized by curvature perturbation is known to
occur, causing a swift increase in the frequency.

\textbf{Keywords: }Pulsar timing array, gravitational wave, inflation,
Hubble tension.
\end{abstract}

\section{Introduction}

The stochastic gravitational wave (GW) signals at nHZ frequencies in the
15-year data set \cite{NANO0,NANO00} and 12.5-year data \cite{NANO1,NANO2},
has recently garnered significant attention. This data has been published by
the North American Nanohertz Observatory for Gravitational Waves (NANOGrav)
collaboration. The GW signal may arise following the power-law distribution
of abundance: $\Omega _{GW}\propto f^{\zeta }$. NANOGrav's observations
indicate that the exponent $\zeta $ lies within the range of $\left(
-1.5,0.5\right) $ at a frequency of $f=5.5nHz$. In response to the NANOGrav
results, a series of recent gravitational wave models, \cite{MNS0,MNS1,ABX1}%
, have been proposed. Often the authors linked this signal with cosmic
inflation \cite{JH1,ABX2} and from axion inflation \cite{AAX0}. This signal
has the potential to be construed as an outcome of the stochastic
gravitational wave background emanating from primordial black holes (PBHs)
formed during inflation, \cite{JCAPN1,AAX10,AX3,BN1,BN2,BN3}. Conversely,
alternative studies propose that cosmic strings might serve as the origin of
these waves, as indicated in \cite{AAX1}.\newline
The authors in \cite{PRL1} explore gravitational wave production during the
end of inflation (preheating). This process generates significant
inhomogeneities, leading to a stochastic background of gravitational waves
within the comoving Hubble horizon at the end of inflation. Importantly,
their results emphasize that the current amplitude of these gravitational
waves is independent of the inflationary energy scale. The analysis focuses
on a specific model with an inflationary energy scale of approximately $%
10^{9}GeV$. This discovery highlights the potential for a new observational
perspective on inflationary physics and encourages further investigations
into stochastic gravitational wave backgrounds across the Hz to GHz
frequency range, characterized by an amplitude of $\Omega _{GW}h^{2}\sim
10^{-11}$. Moreover, their computational approach has broader applications
for understanding gravitational waves originating from various inhomogeneous
processes in the early universe. The paper \cite{JCAP3} considers a scenario
in which the end of inflation is treated as a sudden event, and it provides
mathematical formulations for essential parameters such as the spectrum,
spectral tilt, and non-Gaussianity. These formulas are then applied to
analyze a minimal extension of the initial hybrid inflation model, shedding
light on the intricate dynamics of the early universe.\newline
The Measurements of the current rate of expansion of the Universe conducted
on a local scale, utilizing observations of Type Ia Supernovae (SNeIa)
through the Hubble Space Telescope (HST), exhibit values that tend to
cluster around $H_{0}\simeq \left( 73.24\pm 1.74\right) km.s^{-1}.Mpc^{-1}$
\cite{L1}. In contrast, calculations of the Hubble constant derived from
measurements of the Cosmic Microwave Background (CMB) indicate a value of
approximately $H_{0}\simeq \left( 67.27\pm 0.66\right) km.s^{-1}.Mpc^{-1}$
\cite{L2}. This introduces a deviation of 4$\sigma $ or more from the local
measurements. One of the more effective resolutions to the Hubble tension
\cite{HTE0,HTE1,HTE2,HTE3,HTE4} involves incorporating an early dark energy
component \cite{PL1} or introducing primordial magnetic fields \cite{PL2}.
Additionally, numerous studies have approached this issue by imposing
constraints.\newline
The purpose of the present paper is to introduce an interpretation of the
Hubble tension from NANOGrav signal at the end of inflation.\newline
This paper is organized as follows. In Sec. 2, we review the stochastic
gravitational wave background from the tensor perturbation and the
gravitational wave spectrum. In Sec. 3, we analyze the Hubble tension in
inflationary epoch. In Sec. 4, the e-folding number, slow roll parameter and
spectral index are studied. In Sec. 5, we study the energy density spectrum
in the end of inflation. We conclude our findings in Sec. 6.

\section{Amplitude of the GW signal}

The action of the gravitational wave sources represented by the tensor
perturbation $h_{ij}$ $(i,j=1,2,3)$ is
\begin{equation}
S=\int d\tau d^{3}x\sqrt{-\delta }\left[ \frac{-\delta ^{\mu \nu }}{64\pi G}%
\partial _{\mu }h_{ij}\partial _{\nu }h^{ij}+\frac{1}{2}\Pi _{ij}h^{ij}%
\right] ,
\end{equation}%
where $G$ is the Newtonian gravitational constant, $\tau $ is the conformal
time and $\Pi _{ij}$ corresponds to an anisotropic stress, encompassing the
impact of a macroscopic magnetic field. Here $\delta _{\mu \nu }\ $in the
Minkowski metric of flat spacetime \cite{REP} $g_{\mu \nu }=\delta _{\mu \nu
}+h_{\mu \nu }$ $(\mu ,\nu =0,1,2,3).$ \newline
The Friedmann-Robertson-Walker (FRW) spacetime \cite{JCAP2}:
\begin{equation}
ds^{2}=\left( ic^{2}dt\right) ^{2}-a^{2}(t)\left( \delta _{ij}+h_{ij}\right)
dx^{i}dx^{j},  \label{a1}
\end{equation}%
where $a\left( t\right) $ is the scale factor with $ad\tau =dt$. The Hubble
parameter is defined as $H=\dot{a}/a$ where $\dot{a}$ signifies the
derivative of $a$ with respect to time. we write the conditions on the
tensor $h_{ij}$ as $\partial ^{i}h_{ij}=h_{\text{ \ }i}^{i}=0$.\newline
We can perform a Fourier transformation of the gravitational wave field by
\cite{PRD1,PRD2,MN1}:\newline
\begin{equation}
h_{ij}\left( \tau ,\vec{x}\right) =\int \frac{d^{3}\vec{k}}{\left( 2\pi
\right) ^{3/2}}\sum_{s=+,\times }\left[ h_{k}\left( \tau \right) \epsilon
_{ij}C\left( \vec{k}\right) e^{i\vec{k}\vec{x}}+cc\right] ,
\end{equation}%
where $k=\left\vert \vec{k}\right\vert =2\pi f$ is the GW wavenumber and $cc$
stands for the complex conjugate term. Here, the $\epsilon _{ij}$ satisfy
the conditions $\delta ^{ij}\epsilon _{ij}=0$. The energy density of the
gravitational waves is expressed as follows \cite{JCAP2}:%
\begin{equation}
\rho _{GW}=\frac{1}{64\pi G}\left\langle \left( \partial _{t}h_{ij}\right)
^{2}+\frac{1}{a^{2}}\left( \nabla h_{ij}\right) ^{2}\right\rangle ,
\label{a2}
\end{equation}%
where the bracket $\left\langle \bullet \right\rangle $ describes the
spatial average. To describe the spectral amplitude of GWs, we employ the
dimensionless parameter $\Omega _{GW}=\rho _{GW}/\rho _{crit}$ where $\rho
_{crit}=\frac{3H_{0}^{2}}{8\pi G}$ . This parameter signifies the energy
density of GWs within each logarithmic frequency interval:
\begin{equation}
\Omega _{GW}\left( f\right) =\frac{1}{\rho _{crit}}\frac{d\rho _{GW}}{d\ln f}%
,  \label{a3}
\end{equation}%
where $H_{0}$ is the current Hubble constant and $f$ is the frequency of the
GW signal. An approximation for the gravitational wave spectrum can be made
by employing the power-law model:%
\begin{equation}
\Omega _{GW}\left( f\right) =\Omega _{GW\ast }\left( \frac{f}{f_{\ast }}%
\right) ^{n_{s}\left( f\right) }\text{,}  \label{a4}
\end{equation}%
where $\Omega _{GW\ast }=\Omega _{GW}\left( f=f_{\ast }\right) $. The
spectral index is%
\begin{equation}
n_{s}\left( f\right) =\left\{
\begin{array}{c}
n_{GW1}\text{ \ \ \ for \ \ }f<f_{\ast } \\
n_{GW2}\text{ \ \ \ \ for \ }f>f_{\ast }%
\end{array}%
\right. .  \label{a5}
\end{equation}%
We notice that for \ $f=f_{\ast }$, we get $n_{GW1}=n_{GW2}$. If $n_{s}=-1$,
we obtain $f\Omega _{GW}=f_{\ast }\Omega _{GW\ast }$.

\section{Hubble tension from GW signal}

Hubble tension from GW signal \cite{AX2,AX3}. The current amplitude of the
gravitational wave signal observed today is given by \cite{JCAP1,AX1}%
\begin{equation}
\Omega _{GW}\left( f\right) =\frac{\Omega _{R,0}}{24}\mathcal{P}_{h}\left(
f\right) ,  \label{b1}
\end{equation}%
where $\Omega _{R,0}\approx 8.6\times 10^{-5}$ is the radiation energy
density today and $\mathcal{P}_{h}\left( f\right) $ is the power spectrum of
tensor fluctuations, which varies with frequency, at the moment of exiting
the cosmic horizon. We can establish a connection between the e-folding
number $N$ and the GW frequency of the signal $f$, such that \cite{JCAP1}:%
\begin{equation}
N=N_{CMB}+\ln \frac{k_{CMB}}{0.002Mpc^{-1}}-44.9-\ln \frac{f}{10^{2}Hz}.
\label{b2}
\end{equation}%
We notice that $N=0$ at the end of inflation. While, in the CMB scale we
have $k_{CMB}=0.002Mpc^{-1}$and $N_{CMB}\sim 50-60$, in this case we get%
\begin{equation}
N=N_{CMB}-44.9-\ln \frac{f}{10^{2}Hz}.  \label{b3}
\end{equation}%
Or equivalently%
\begin{equation}
N=\ln \frac{10^{2}e^{N_{CMB}-44.9}Hz}{f}.  \label{b4}
\end{equation}%
Thus, the e-folding number can be expressed as%
\begin{equation}
N=\ln \frac{f_{\ast }}{f},  \label{b5}
\end{equation}%
where $f_{\ast }=10^{2}e^{\left( N_{CMB}-44.9\right) }Hz$ is the typical
frequency. Based on this relation, it's noteworthy that the state
represented by $f_{\ast }=f$ describes the end of inflation. Conversely, in
order to align the value of $N$ with observational data, we need to take
into account the following condition:
\begin{equation}
f<f_{\ast }.  \label{b7}
\end{equation}%
This condition is comparable with $n_{s}=n_{GW1}$ (\ref{a5}). On the other
hand, the equations (\ref{a4}) and (\ref{b5}) take this form,%
\begin{equation}
\Omega _{GW}\left( f\right) =\Omega _{GW\ast }e^{n_{s}\left( f\right) N\text{
}}.  \label{b8}
\end{equation}%
At the end of inflation, we find $\Omega _{GW}\left( f\right) \approx \Omega
_{GW\ast }$. The Planck CMB data implies that $n_{s}\approx 0.965\pm 0.004$
\cite{L2}. It is easy to see from $N_{CMB}\sim 50$ that%
\begin{equation}
\Omega _{GW}\left( f\right) =\Omega _{GW\ast }e^{48.25\pm 0.2\text{ }}.
\label{ns1}
\end{equation}%
For $\Omega _{GW}\approx 9.3_{-4.0}^{+5.8}\times 10^{-9}$ \cite{NANO0,NANO00}
we have the critical density of GWs as follows $\Omega _{GW\ast }\sim
10^{-29}-10^{-30}$. The number of e-folds can be given by \cite{Ann1,Ann2}:
\begin{equation}
N=\frac{1}{\epsilon _{H}}\int_{H}^{H_{end}}d\ln H^{\prime }=\frac{1}{%
\epsilon _{H}}\ln \frac{H_{end}}{H}.  \label{b9}
\end{equation}%
Here, $\epsilon _{H}=-\frac{\dot{H}}{H^{2}}\ll 1$ is the slow roll
parameter. This relationship corresponds with observations data, because if
we take $H=H_{end}$ we find the value of $N$ at the end of inflation ($N=0$%
). We can additionally calculate the $H$ concerning $f$ using the following
approach:%
\begin{equation}
H\left( f\right) =H_{end}\left( \frac{f}{f_{\ast }}\right) ^{\epsilon _{H}}.
\label{b10}
\end{equation}%
In the scenario where $f_{\ast }=f$, we arrive at $H\left( f\right) =H_{end}$%
. Conversely, in the case of (\ref{b7}), we achieve%
\begin{equation}
H\left( f\right) <H_{end}.  \label{b11}
\end{equation}%
We recall that the assessment of the present rate of the Universe's
expansion yields $H_{0}\simeq \left( 73.24\pm 1.74\right) km.s^{-1}.Mpc^{-1}$
\cite{L1}. Conversely, measurements of the Hubble constant based on the CMB
result is $H_{0}\simeq \left( 67.27\pm 0.66\right) km.s^{-1}.Mpc^{-1}$ \cite%
{L2}. The Hubble tension's behavior reveals a diminishing value of $H$ as
time elapses. This shows that according to (\ref{b11}) the Hubble parameter $%
H\left( f\right) $ characterizes the inflationary epoch. This is underscored
by the fact that $H_{end}$ signifies the end of inflation. Consequently, the
frequencies $f$ detected by the NANOGrav collaboration are indicative of the
inflationary era.

\section{Inflationary scenario from e-folding number}

Thereafter we want to seek the relationship between the current amplitude of
the gravitational wave signal and frequency. We recall that $\dot{N}=-H$
\cite{L15}, i.e. $\dot{N}=-\frac{\dot{f}}{f}$ (\ref{b5}). Consequently, this
leads to
\begin{equation}
H=\frac{\dot{f}}{f}.  \label{c10}
\end{equation}%
This result implies that the Hubble parameter represents the logarithmic
variation of GW frequency. If we compare (\ref{c10}) with the definition of
the Hubble parameter $H=\frac{\dot{a}}{a}$, we find $f=af_{0}$ with $f_{0}$
is a constant with frequency unit. This proportionality between the scale
factor and frequency can establish a link between the Hubble tension problem
and the NANOGrav 15-yr result. The slow roll parameter $\epsilon _{H}=-\frac{%
\dot{H}}{H^{2}}$ can be written as
\begin{equation}
\epsilon _{H}=1-\frac{\ddot{f}f}{\dot{f}^{2}}.  \label{c11}
\end{equation}%
Introducing these parameters%
\begin{equation}
\sigma =\frac{\ddot{f}}{\dot{f}^{2}}\text{, \ }\lambda =\frac{\dddot{f}}{%
\ddot{f}}\frac{1}{\dot{f}}.
\end{equation}%
From this model, the slow-roll\ parameter $\eta =\epsilon _{H}-\frac{\dot{%
\epsilon}_{H}}{2H\epsilon _{H}}$ \cite{L15} is given by%
\begin{equation}
\eta =1-\sigma f-\frac{f}{2\left( 1-\sigma f\right) }\left( 2\sigma
^{2}f-\lambda \sigma f-\sigma \right) .  \label{c12}
\end{equation}%
In the inflationary scenario, the $\epsilon _{H}$ grows as well $d\epsilon
_{H}/dN=2\epsilon _{H}\left( \eta -\epsilon _{H}\right) $. The spectral
index of perturbations is equal to \cite{L15}:
\begin{equation}
n_{s}=1-6\epsilon _{H}+2\eta .  \label{c14}
\end{equation}%
Thus, the spectral index (see Appendix A) can be written as%
\begin{equation}
n_{s}=-3+4\sigma f-\frac{f}{1-\sigma f}\left( 2\sigma ^{2}f-\lambda \sigma
f-\sigma \right) .  \label{c15}
\end{equation}%
If $\lambda =2\sigma $, we can estimate the spectral index by the ratio $%
n_{s}\approx -3+4\sigma f+\frac{\sigma f}{1-\sigma f}$. If $f=0$ we obtain $%
n_{s}=-3$. From (\ref{b8}) and (\ref{c15}) we get%
\begin{equation}
\Omega _{GW}\left( f\right) =\Omega _{GW\ast }\exp N(-3+4\sigma f-\frac{%
\sigma f}{1-\sigma f}\left( 2\sigma f-\lambda f-1\right) ),  \label{c16}
\end{equation}%
where $\Omega _{GW\ast }$ is the amplitude at the typical frequency $f_{\ast
}$. The amplitude at $f=0$ is given by $\Omega _{GW}\left( f\right) =\Omega
_{GW\ast }e^{-6N\text{ }}$.

\section{Stochastic GW signals at preheating}

\begin{figure}[H]
\centering\includegraphics[width=10cm]{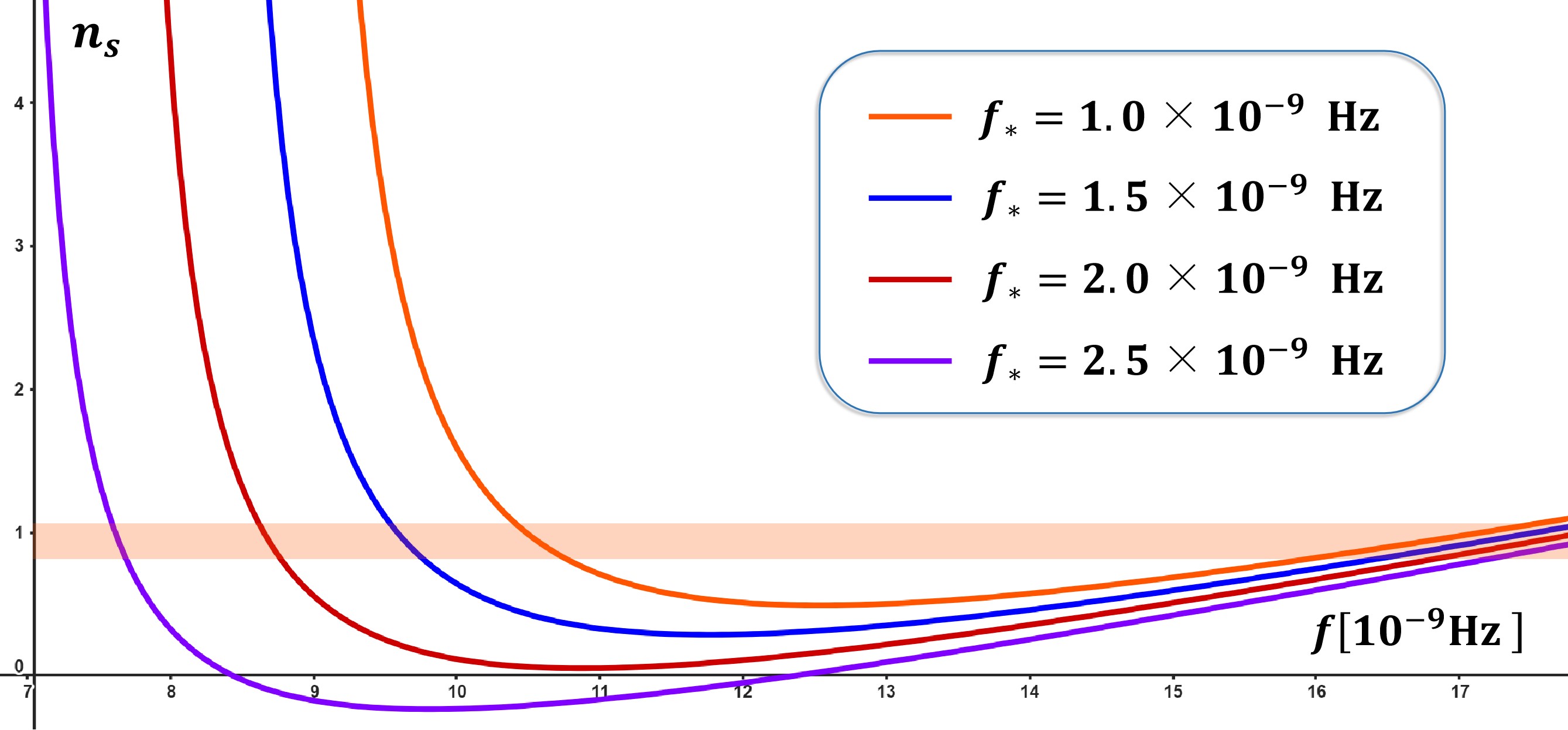}
\caption{ Evolution of model parameters $n_{s}$ (\protect\ref{d7}) for $%
\protect\sigma =5\times 10^{7}$ and $nHz\leq f_{\ast }\leq 2.5nHz.$}
\label{F1}
\end{figure}

\begin{figure}[H]
\centering\includegraphics[width=10cm]{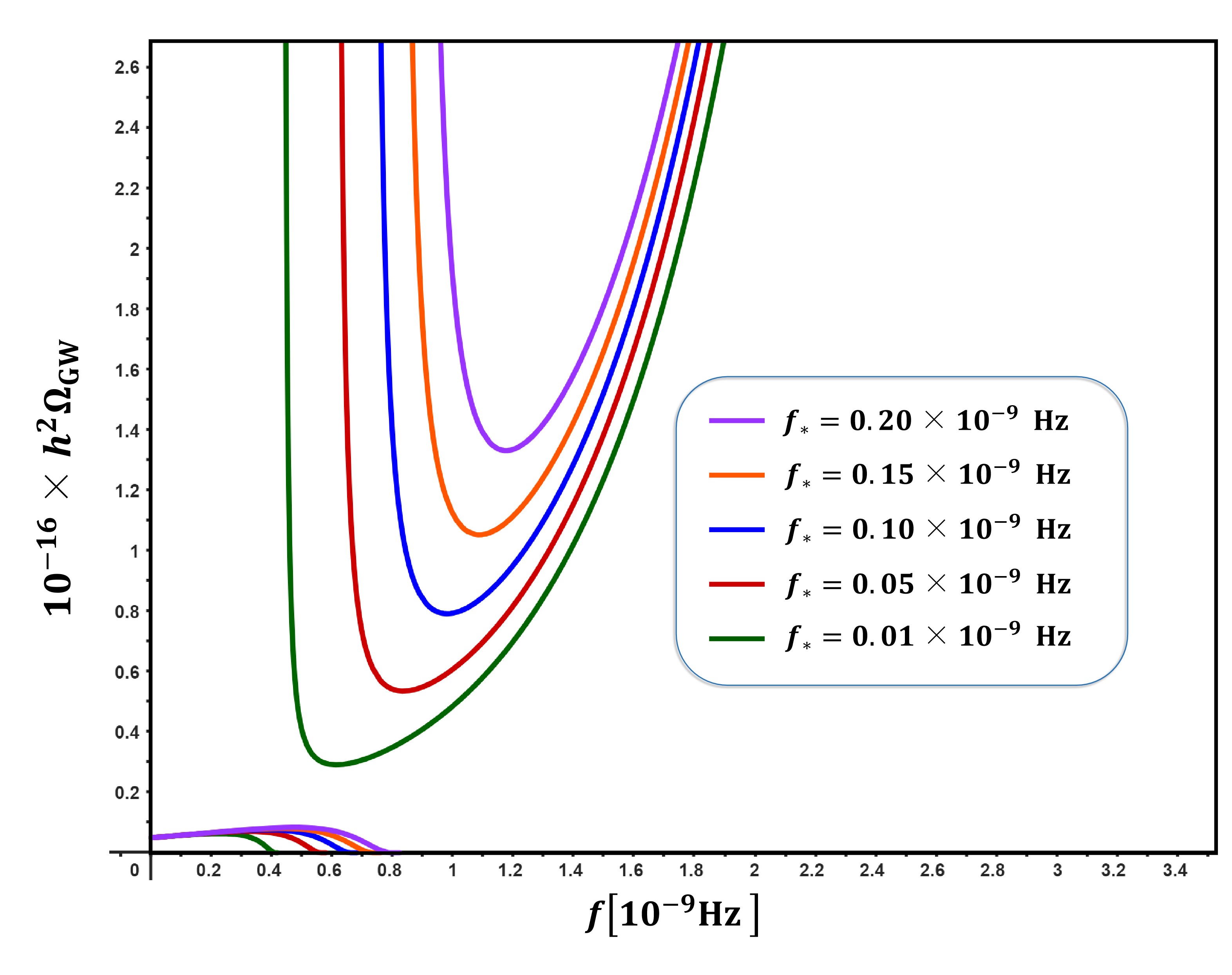}
\caption{Evolution of energy density spectrum $\Omega _{GW}h^{2}$ (\protect
\ref{d8}) for $\protect\sigma =5\times 10^{7}$ and $N=1$ with $nHz\leq
f_{\ast }\leq 2.5nHz.$ }
\label{F2}
\end{figure}

\begin{figure}[H]
\centering\includegraphics[width=10cm]{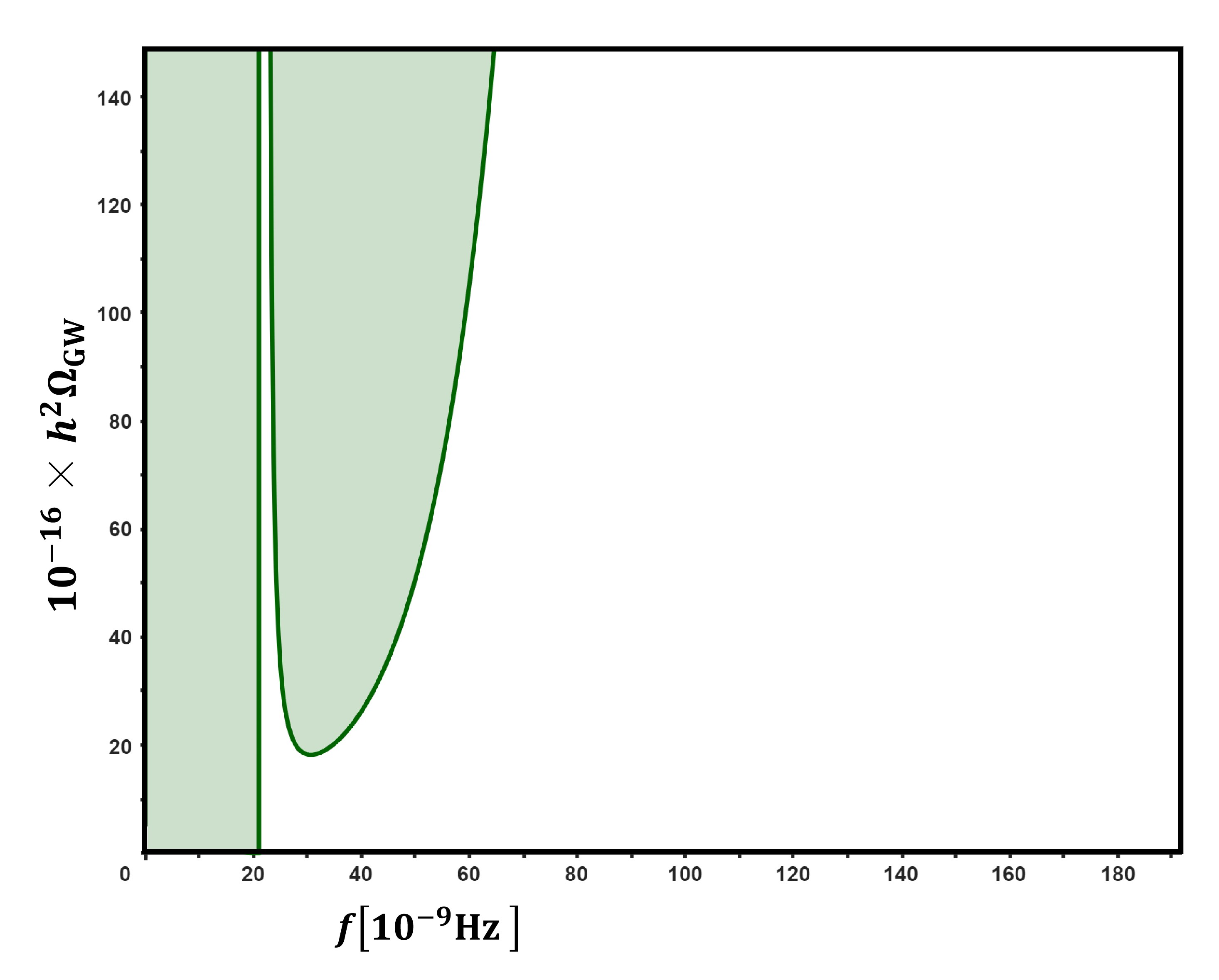}
\caption{Region of the gravitational wave spectrum for $\protect\sigma %
=10^{7}$, $\Omega _{GW\ast }\sim 10^{-14}$, $N=2$ and $f_{\ast }=0.5nHz.$ }
\label{F3}
\end{figure}

\begin{figure}[H]
\centering\includegraphics[width=10cm]{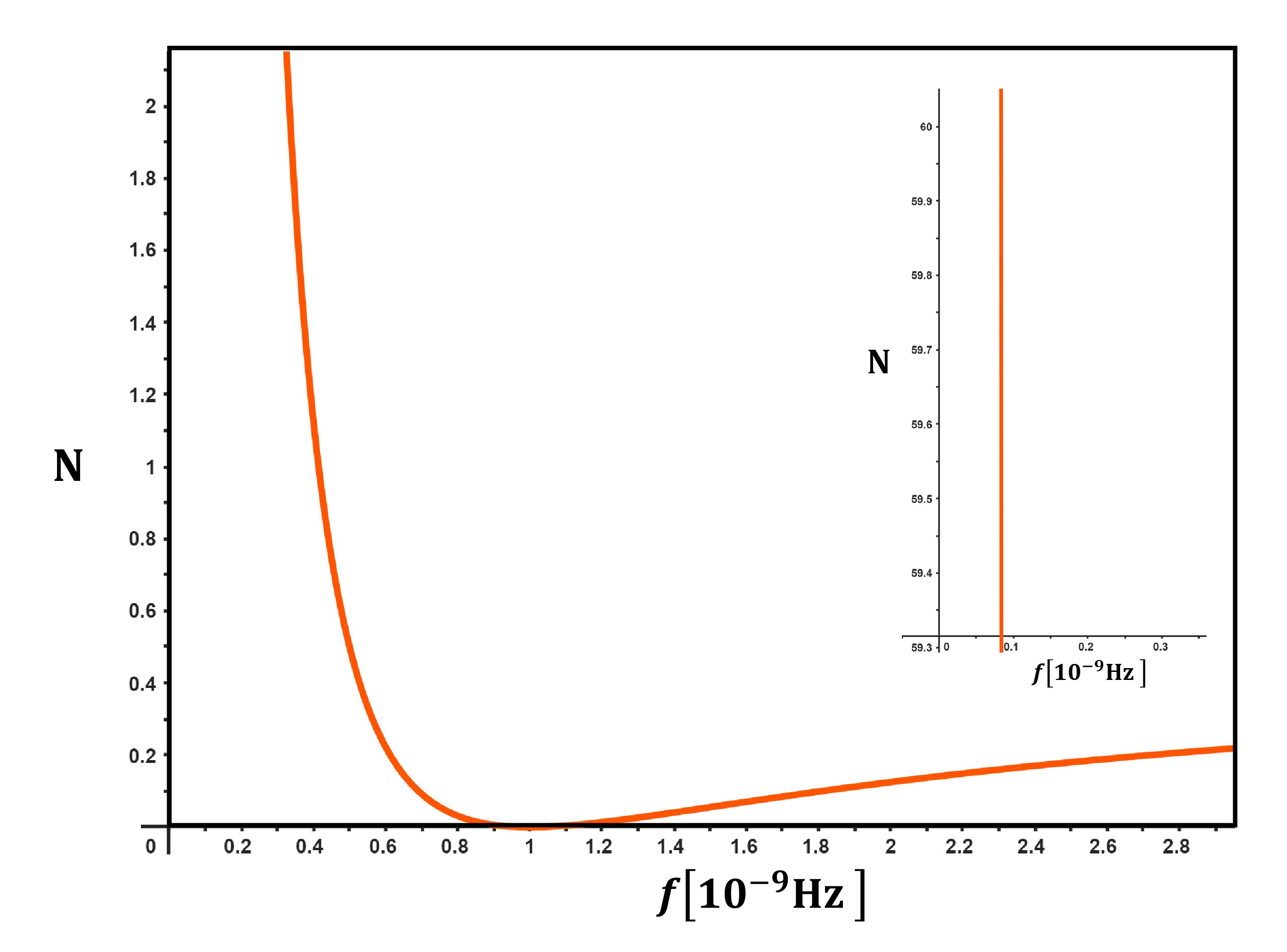}
\caption{$N$ curve as a function of $f$, such as $\protect\sigma =1.5\times
10^{7}s$ and $\protect\zeta =-0.5.$ }
\label{F4}
\end{figure}

Thereafter, we want to study the spectral index from the relationship (\ref%
{b10}) instead of (\ref{c10}): $H\left( f\right) =H_{end}\left( \frac{f}{%
f_{\ast }}\right) ^{\epsilon _{H}}.$ We calculate the derivative of this
parameter with respect to time (see Appendix A), and we determine%
\begin{equation}
\frac{\dot{H}\left( f\right) }{H\left( f\right) }=\dot{\epsilon}_{H}\ln
\left( \frac{f}{f_{\ast }}\right) +\epsilon _{H}\frac{\dot{f}}{f}.
\label{d2}
\end{equation}%
Using $\epsilon _{H}=-\frac{\dot{H}}{H^{2}}$, we can relate the slow roll
parameter $\epsilon _{H}$ to GW frequency as%
\begin{equation}
-\frac{H\epsilon _{H}}{\dot{\epsilon}_{H}}=\ln \left( \frac{f}{f_{\ast }}%
\right) +\frac{\epsilon _{H}\dot{f}}{\dot{\epsilon}_{H}f},  \label{d3}
\end{equation}%
and hence it can be written as%
\begin{equation}
\epsilon _{H}-\frac{\dot{\epsilon}_{H}}{2H\epsilon _{H}}=\epsilon _{H}+\frac{%
1}{2}\left[ \ln \left( \frac{f}{f_{\ast }}\right) +\frac{\epsilon _{H}\dot{f}%
}{\dot{\epsilon}_{H}f}\right] ^{-1}.  \label{d4}
\end{equation}%
Therefore, the slow-roll\ parameter $\eta =\epsilon _{H}-\frac{\dot{\epsilon}%
_{H}}{2H\epsilon _{H}}$ \cite{L15} is given by $\eta =\epsilon _{H}+\frac{1}{%
2}\left[ \frac{\epsilon _{H}\dot{f}}{\dot{\epsilon}_{H}f}+\ln \left( \frac{f%
}{f_{\ast }}\right) \right] ^{-1}$. Thus, the spectral index $%
n_{s}=1-6\epsilon _{H}+2\eta $ can be written as%
\begin{equation}
n_{s}=1-4\epsilon _{H}+\left[ \frac{\epsilon _{H}\dot{f}}{\dot{\epsilon}_{H}f%
}+\ln \left( \frac{f}{f_{\ast }}\right) \right] ^{-1}.  \label{d6}
\end{equation}%
If $\sigma f=1$, one obtain $n_{s}=1+\ln ^{-1}\left( \frac{f}{f_{\ast }}%
\right) $, i.e. $\Omega _{GW}\left( f\right) =\Omega _{GW\ast }e^{\left(
1+\ln ^{-1}\left( \frac{f}{f_{\ast }}\right) \right) N\text{ }}$. Next, we
consider $\sigma =\frac{\ddot{f}}{\dot{f}^{2}}$\ to be a constant and we
obtain $\epsilon _{H}=1-\sigma f$, this yields $\frac{\dot{\epsilon}_{H}}{%
\epsilon _{H}}=\frac{-\sigma \dot{f}}{1-\sigma f}$. The $n_{s}$\ with
respect to $f$\ is given by%
\begin{equation}
n_{s}=-3+4\sigma f+\left[ -\frac{1-\sigma f}{\sigma f}+\ln \left( \frac{f}{%
f_{\ast }}\right) \right] ^{-1},  \label{d7}
\end{equation}%
and the energy density spectrum can be written as%
\begin{equation}
\Omega _{GW}\left( f\right) =\Omega _{GW\ast }\exp N\left( -3+4\sigma f+
\left[ -\frac{1-\sigma f}{\sigma f}+\ln \left( \frac{f}{f_{\ast }}\right) %
\right] ^{-1}\right) .  \label{d8}
\end{equation}%
In Figure (\ref{F1}) we plot the $n_{s}$ as a function of the frequency $f$.
In this illustration, it's evident that the curves representing the function
$n_{s}(f)$ intersect with the range of observed values. While, in Figures (%
\ref{F2})-(\ref{F3}) we plot the energy density spectrum $\Omega _{GW}h^{2}$
as it varies with frequency $f$.
In Figure (\ref{F3}), we observe two distinct regions delineating the GW
frequency characteristics of the early universe. The interval $0nHz\leq
f\leq 20nHz$ corresponds to the outcomes associated with the CMB. Notably,
the interval $f\sim 25nHz-65nHz$ characterizes the dataset generated by the
NANOGrav 15-yr result at the (near) end of inflation $N\sim 2$. These
results directly match the observed data \cite{NANO0,NANO00,AX1}. This
constraint on $f$ has significant implications for the energy density
spectrum of gravitational waves, $\Omega _{GW}h^{2}$, at PTA frequencies. A
vanishing GW frequency, which would imply a nearly scale-invariant GW
spectrum, corresponds to the first instance of the universe. The study
suggests that there exists an intermediate range of frequencies between
those of CMB and PTA, which lacks energy density spectrum. components.%
\newline
We consider that $\sigma f\ll 1$ and $-3+4\sigma f-\frac{\sigma f}{1-\sigma f%
}=0$, thus, (\ref{d7}) can be written as $n_{s}\approx -\left( \frac{\sigma f%
}{1-\sigma f}\right) ^{2}\ln \left( \frac{f}{f_{\ast }}\right) $. The energy
density spectrum is estimated as%
\begin{equation}
\Omega _{GW}\left( f\right) h^{2}=\Omega _{GW\ast }h^{2}\left( \frac{f}{%
f_{\ast }}\right) ^{-\left( \frac{\sigma f}{1-\sigma f}\right) ^{2}N\text{ }%
}.  \label{d9}
\end{equation}%
If we use $k=2\pi f$ with $k_{\ast }=2\pi f_{\ast }$ is the characteristic
scale, we find that $\Omega _{GW}\left( f\right) =\Omega _{GW\ast }\exp
\left( -\ln \left( \frac{k}{k_{\ast }}\right) \left( \frac{\sigma f}{%
1-\sigma f}\right) ^{2}N\right) $, which exactly corresponds to the
primordial power spectrum of curvature perturbations $\mathcal{P}_{\mathcal{R%
}}\left( f\right) $ \cite{AX4,AX5}, also it checked equation (\ref{b1}). The
relation (\ref{d9}) is in good agreement with (\ref{a4}) and the NANOGrav's
observations, suggesting that $\Omega _{GW}\propto f^{\zeta }$ with $\zeta
\in \left( -1.5,0.5\right) $ at a frequency of $f=5.5nHz$. Combining this
with (\ref{d9}) we get%
\begin{equation}
\zeta =-\left( \frac{\sigma f}{1-\sigma f}\right) ^{2}N<0.  \label{d10}
\end{equation}%
In the scenario where $\sigma \sim 10^{7}$, as illustrated in Figures (\ref%
{F2})-(\ref{F3}), we determine that $\zeta =-0.143876N<0$. This conclusion
subsequently implies the following condition: $0\leq N\leq 10.425$.
Furthermore, this range ensures that $\zeta $ remains greater than or equal
to $-1.5$. In the tables below (\ref{T1}), we analyze numerical
representations of the relationship (\ref{d10}).
\begin{table}[H]
\begin{center}
\begin{equation*}
\begin{tabular}{ccc}
\hline
$N$ & $\zeta $ & $\sigma \left[ s\right] $ \\ \hline\hline
$60$ & $-1.5$ & $2.482\times 10^{7}$ \\
$60$ & $-0.5$ & $1.520\times 10^{7}$ \\
$60$ & $-10^{-6}$ & $2.346\times 10^{4}$ \\
$50$ & $-1.5$ & $2.684\times 10^{7}$ \\
$50$ & $-0.5$ & $1.652\times 10^{7}$ \\
$2$ & $-0.5$ & $6.060\times 10^{7}$ \\ \hline
\end{tabular}%
\end{equation*}%
\end{center}
\caption{The stochastic gravitational wave background based on the value of
the Hubble constant for $f=5.5nHz$.}
\label{T1}
\end{table}
Based on these results, it is worth noting that the values of $\zeta $
precisely satisfy the data conditions indicating that the relationship (\ref%
{b10}) aligns well with the observational data. The range $0\leq N\leq
10.425 $ signifies the epoch marking the end of inflation. During this
period, there is the emission of stochastic gravitational waves, which are
then detected by the NANOGrav collaboration. From the relationship (\ref{d10}%
) we deduce $N\left( f\right) =\left( \frac{1-\sigma f}{\sigma f}\right)
^{2}\left\vert \zeta \right\vert $ and $f=\frac{1}{\sigma \left( 1+\sqrt{%
\frac{N}{\left\vert \zeta \right\vert }}\right) }$. From this, we plot the
curve of $N$ as a function of $f$ in figure (\ref{F4}).
In Figure (\ref{F4}), we observe that when considering the CMB with $30\leq
N\leq 60$ \cite{MNS3}, the frequency of the GW signal $f$ remains
consistently close to the value $f=0.08nHz$. However, as we approach the end
of inflation, the range of $N\leq 1$ values encompasses two distinct
frequency values. Finally, at the precise end of inflation ($N=0$), the
frequency settles at a singular value, which is equal to $f=1nHz$. The sharp
increase in frequency observed towards the end of the inflationary period
indicates a significant cosmic variation. This change is due to the
curvature perturbation \cite{JCAP3}.\newline
The state of the universe at the the end of inflation is known as the
preheating \cite{PRL1}. Typically, following the end of inflation, our
Universe will transition into a radiation-dominated era, during which
perturbations will once again enter the observable horizon.

\section{Conclusion}

In summary, the study has explored the stochastic gravitational wave
produced during the end of inflation. In this context, it has been
established that the Hubble parameter is linked to the logarithmic variation
of GW frequency. By analyzing the frequencies detected by the NANOGrav
collaboration in relation to the end of inflation marked by $H_{end}$, we
have unveiled a significant connection. This connection suggests a
logarithmic relationship between the Hubble parameter and gravitational wave
frequency, represented as $f\propto a$, where $a$ represents the scale
factor. This proportional link between scale factor and frequency could shed
light on the Hubble tension problem and its connection to the NANOGrav 15-yr
results. Furthermore, we present plots of the spectral index $n_{s}$ as a
function of frequency $f$, showing how $n_{s}\left( f\right) $ intersects
with observed values. We have also explored the energy density spectrum $%
\Omega _{GW}h^{2}$ in relation to frequency, revealing two distinct regions
in Figure (\ref{F3}), corresponding to GW frequencies associated with the
early universe and those generated by NANOGrav 15-yr results at the end of
inflation. These results align with observed data, emphasizing the
importance of frequency constraints for the energy density spectrum of
gravitational waves at PTA frequencies. Notably, an absence of energy
density spectrum components is identified in an intermediate frequency
range, bridging the CMB and PTA domains. In Figure (\ref{F4}), we have
observed the behavior of GW signal frequency concerning the CMB for
different inflationary parameters. Remarkably, as we approach the end of
inflation ($N\leq 1$), two distinct frequency values emerge. At the precise
end of inflation ($N=0$), the frequency stabilizes at a singular value, $%
f=1nHz$. This study proposes that primordial universe scenarios, which rely
on the emission of the stochastic gravitational waves at the end of
inflation, can offer a new period of the emission of this stochastic
gravitational wave background. This could result in a greater likelihood of
observing distinct signals in PTA observations.

\section{Appendix A}

\textbf{1- Inflationary scenario from e-folding number}\newline
From $\dot{N}=-H$ \ we have $\dot{N}=-\frac{\dot{f}}{f}=-H$, which leads to $%
H=\frac{\dot{f}}{f}$. From this we get%
\begin{equation}
\dot{H}=\frac{\ddot{f}f-\dot{f}^{2}}{f^{2}}.
\end{equation}%
We employ the constraint $\epsilon _{H}=-\frac{\dot{H}}{H^{2}}=\frac{\dot{f}%
^{2}-\ddot{f}f}{\dot{f}^{2}}$derived from $\epsilon _{H}=-\frac{\dot{H}}{%
H^{2}}$ and we find
\begin{equation}
\epsilon _{H}=1-\frac{\ddot{f}f}{\dot{f}^{2}}=1-\sigma f,
\end{equation}%
where $\sigma =\frac{\ddot{f}}{\dot{f}^{2}}$ and $\lambda =\frac{\dddot{f}}{%
\ddot{f}}\frac{1}{\dot{f}}$. Using $2H\epsilon _{H}=2\frac{\dot{f}^{2}-\ddot{%
f}f}{\dot{f}f}$ and $\frac{1}{2H\epsilon _{H}}=\frac{\dot{f}f}{2\left( \dot{f%
}^{2}-\ddot{f}f\right) }$ we can calculate the slow-roll\ parameter $\eta
=\epsilon _{H}-\frac{\dot{\epsilon}_{H}}{2H\epsilon _{H}}$. We have%
\begin{equation}
\dot{\epsilon}_{H}=2\left( \frac{\ddot{f}}{\dot{f}}\right) ^{2}\frac{f}{\dot{%
f}}-\frac{\dddot{f}}{\dot{f}^{2}}f-\frac{\ddot{f}}{\dot{f}},
\end{equation}%
which leads to%
\begin{equation}
\eta =1-\frac{\ddot{f}f}{\dot{f}^{2}}-\frac{f}{2\left( 1-\frac{\ddot{f}f}{%
\dot{f}^{2}}\right) }\left( 2\left( \frac{\ddot{f}}{\dot{f}}\right) ^{2}%
\frac{f}{\dot{f}^{2}}-\frac{\dddot{f}}{\dot{f}}\frac{f}{\dot{f}^{2}}-\frac{%
\ddot{f}}{\dot{f}^{2}}\right) .
\end{equation}%
So we can relate the slow-roll\ parameter to frequency of the GW signal $f$
as
\begin{equation}
\eta =1-\sigma f-\frac{f}{2\left( 1-\sigma f\right) }\left( 2\sigma
^{2}f-\lambda \sigma f-\sigma \right) .
\end{equation}%
We can additionally calculate the spectral index using the following
approach: the spectral index $n_{s}=1-6\epsilon _{H}+2\eta $ is
\begin{equation}
n_{s}=1-6+6\sigma f+2-2\sigma f-\frac{f}{\left( 1-\sigma f\right) }\left(
2\sigma ^{2}f-\lambda \sigma f-\sigma \right) .
\end{equation}%
So, we find equation (\ref{c14}):%
\begin{equation}
n_{s}=-3+4\sigma f-\frac{f}{\left( 1-\sigma f\right) }\left( 2\sigma
^{2}f-\lambda \sigma f-\sigma \right) .
\end{equation}

\textbf{2-} \textbf{Inflationary scenario from Hubble parameter}\newline
From the relationship (\ref{b10}) $H\left( f\right) =H_{end}\left( \frac{f}{%
f_{\ast }}\right) ^{\epsilon _{H}}$ we have%
\begin{equation}
\dot{H}\left( f\right) =\frac{d}{dt}\left( \epsilon _{H}\ln \left( \frac{f}{%
f_{\ast }}\right) \right) H\left( f\right) ,
\end{equation}%
where%
\begin{equation*}
\frac{d}{dt}\left( \epsilon _{H}\ln \left( \frac{f}{f_{\ast }}\right)
\right) =\dot{\epsilon}_{H}\ln \left( \frac{f}{f_{\ast }}\right) +\epsilon
_{H}\frac{\dot{f}}{f},
\end{equation*}%
which leads to%
\begin{equation}
\frac{\dot{H}\left( f\right) }{H\left( f\right) }=\left( \dot{\epsilon}%
_{H}\ln \left( \frac{f}{f_{\ast }}\right) +\epsilon _{H}\frac{\dot{f}}{f}%
\right) .
\end{equation}%
Using $\epsilon _{H}=-\frac{\dot{H}}{H^{2}}$, so we find equation (\ref{d2})
$-H\epsilon _{H}=\dot{\epsilon}_{H}\ln \left( \frac{f}{f_{\ast }}\right)
+\epsilon _{H}\frac{\dot{f}}{f}.$

\end{document}